\documentclass[sigconf]{acmart}
\copyrightyear{2020} 
\acmYear{2020} 
\setcopyright{acmlicensed}
\acmConference[ASIA CCS '20]{Proceedings of the 15th ACM Asia Conference on Computer and Communications Security}{October 5--9, 2020}{Taipei, Taiwan}
\acmBooktitle{Proceedings of the 15th ACM Asia Conference on Computer and Communications Security (ASIA CCS '20), October 5--9, 2020, Taipei, Taiwan}
\acmPrice{15.00}
\acmDOI{10.1145/3320269.3384759}
\acmISBN{978-1-4503-6750-9/20/06}

\usepackage{booktabs}   
\usepackage{subcaption}
\usepackage[justification=centering,skip=0pt]{caption}
\usepackage[utf8]{inputenc}
\usepackage[T1]{fontenc}
\usepackage{amssymb}
\usepackage{amsmath}
\usepackage{graphicx}
\usepackage{multicol}
\usepackage{color}
\usepackage{listings}
\usepackage{fancyvrb}
\usepackage{mathtools}
\usepackage{algorithm}
\usepackage[noend]{algpseudocode}
\usepackage{semantic}

\let\counterwithin\relax
\usepackage{chngcntr}
\usepackage{url}
\usepackage{paralist}
\usepackage{enumitem}
\usepackage{framed}
\usepackage{mdframed}
\usepackage{mathpartir}
\usepackage{tabularx}
\usepackage{balance}

\counterwithin*{paragraph}{subsection}

\begin{document}
\fancyhead{}

\newcommand{\framework}{Estrela}
\newcommand{\user}{\mathbb{U}}
\newcommand{\policy}{\mathbb{P}}
\newcommand{\api}{\mathbb{A}}

\title{Contextual and Granular Policy Enforcement in \\Database-backed Applications} 

\author{Abhishek Bichhawat}
\affiliation{%
  \institution{Carnegie Mellon University}
  \city{Pittsburgh}
  \state{PA}
  \country{USA}
}

\author{Matt Fredrikson}
\affiliation{%
  \institution{Carnegie Mellon University}
  \city{Pittsburgh}
  \state{PA}
  \country{USA}
}

\author{Jean Yang}
\affiliation{%
  \institution{Carnegie Mellon University}
  \city{Pittsburgh}
  \state{PA}
  \country{USA}
}

\author{Akash Trehan}
\authornote{Work done while interning at Carnegie Mellon University.}
\affiliation{%
  \institution{Microsoft Vancouver}
  \city{Vancouver}
  \state{British Columbia}
  \country{Canada}
}

\renewcommand{\shortauthors}{Bichhawat, et al.}

\begin{abstract}
Database-backed applications rely on inlined policy checks to process
users' private and confidential data in a policy-compliant
manner as traditional database access control mechanisms cannot
enforce complex policies. However, application bugs due to missed
checks are common in such applications, which result in data
breaches. While separating policy from code is a natural solution,
many data protection policies specify restrictions based
on the context in which data is accessed and how the data is
used. Enforcing these restrictions automatically presents significant
challenges, as the information needed to determine context requires a
tight coupling between policy enforcement and an application's
implementation.   

We present {\framework}, a framework for enforcing contextual and
granular data access policies. Working from the observation that API
endpoints can be associated with salient 
contextual information in most database-backed applications,
{\framework} allows developers to specify \emph{API-specific}
restrictions on data access and use. {\framework} provides a clean
separation between policy specification and the application's
implementation, which facilitates easier 
auditing and maintenance of policies. Policies in {\framework} consist
of pre-evaluation and post-evaluation conditions, which provide the
means to modulate database access before a query is issued, and to
impose finer-grained constraints on information release after the
evaluation of query, respectively. We build a prototype of
{\framework} and apply it to retrofit several real world applications
(from 1000-80k LOC) to enforce different contextual policies. Our
evaluation shows that {\framework} can enforce policies with minimal
overheads.  
\end{abstract}



\maketitle

\lstset{
basicstyle=\sffamily,
columns=fullflexible,
breaklines=true,
language=python,
postbreak=\mbox{\space},
showstringspaces=false,
numbers=none,
tabsize=1,
captionpos=b,
escapeinside={(*}{*)},
numbersep=3pt,
numberstyle=\tiny,
literate={-}{\textendash}1 {--}{\textemdash}2,
keywordstyle=\sffamily,
}

\section{Introduction}
\label{sec:intro}

Modern systems collect a plethora of personal user
information~\cite{pricon}, but their growing complexity makes it 
increasingly difficult to ensure compliance with data protection
policies, and to avoid widely-publicized data
breaches~\cite{databreach}.  Application-wide policy compliance checks 
are a widely-used way to address these concerns, but ensuring correct
enforcement across component boundaries in applications is
particularly challenging because it requires coordinating such inline 
checks across different parts of the application. Thus, enforcing
compliance in applications and updating them with new regulations, as
they are introduced, is time-consuming, expensive, and error-prone.  

The policies that regulations often require, and that these
applications need to enforce, include more than just access conditions
--- they specify how the data can flow and be used by different users,
and are \emph{contextual} or context-dependent.  
Some other policies might require \emph{declassification}, revealing 
partial information about sensitive data in certain cases. Both
contextual and information release policies are 
prevalent in major privacy laws~\cite{gdpr,hipaa,glb} and
organizational policies~\cite{bing,google,fb}.  

\subsection{Motivation}
We illustrate the need for supporting contextual and release policies
using the example of the Health Insurance Portability and
Accountability Act (HIPAA) privacy rule~\cite{hipaa}.

\paragraph{H1}
While some sections of HIPAA restrict the use or disclosure of
protected health information (PHI) for marketing or research, other
sections allow disclosure if the patient is in need of emergency
treatment. Such \emph{contextual} policies depend on the setting and
circumstances of data access and use, i.e., sensitive data may need to
be retrieved differently, or disallowed, depending on the conditions
under which access is needed (e.g., purpose of access).
As it is essential that in case of an emergency a doctor in
the emergency  department must be allowed access to the PHI of the
patient, normal authorization rules might not apply.

\paragraph{H2}
HIPAA also includes declassification policies --- for instance,
hospitals are allowed to release aggregate information like the total
number of patients diagnosed with a particular disease to be released
as it is not PHI (although derived from it), but the individual
details of affected patients may not be revealed. For declassification
policies, the same set of data accessed individually and together must
be subject to different policies. 

\medskip
The main challenge in specifying and enforcing such policies
lies in the fact that doing so may require detailed information about
the application's internal state and architecture, as well as the
underlying database (whether the data is accessed for emergency or
regular treatment, or is it a part of some aggregate information or is
individually identifiable).

There are numerous existing options for enforcing data protection
policies in database-backed applications. Most enterprise database
systems incorporate role-based fine-grained access control
mechanisms~\cite{oraclevpd, ibmdb2, sepostgresql} that can enforce
policies that prohibit specified users and groups from accessing
certain tables, rows, columns or cells, or provide policy-specific
views of the database to different roles. However, because these
policies cannot refer to information about the applications that use
the database, they are not sufficient for enforcing the contextual
policies needed by many applications, and may introduce performance
overheads when propagating updates for simpler
policies~\cite{mehta2017}.

Several instances of prior work~\cite{mehta2017, upadhyaya2015,
  LeFevre2004, Wang2007, Rizvi2004, Byun2008, Stonebraker1974} have
explored dynamic query-transformation or query-rewriting methods to 
enforce different sorts of policies, but these approaches do not
immediately address the challenge of managing context and application
state, and do not support partial release policies that require
modifications to query responses. For instance, in the emergency
access example (\emph{H1}), some of the proposed
approaches~\cite{LeFevre2004, Byun2008, Kabir2011} include additional
conditions like purpose with the traditional access control mechanisms
where the user specifies the purpose of access additionally, which is
then used to determine access. This, however, requires the electronic
health record systems (EHRs) to \emph{trust} the user requesting
access who may not always be benign~\cite{Jiang2019}. Other EHRs might
provide a single access point that \emph{always} allow access to the
data, and use after-the-fact auditing for ensuring compliance. 


\subsection{Goals, Insights and Contributions}
A proper solution to enforce policies in database-backed applications 
to ensure compliance with data protection laws and regulations should
meet the following criteria in addition to supporting basic cell-,
row- and column-level access policies: 
\begin{enumerate}
\item \emph{Contextual data flow policies:}  
  The enforcement mechanism should be able to enforce different
  policies based on the context of data access or the current state of
  the application.
  
\item \emph{Granular policies for complete mediation:}
  Apart from supporting cell-, row- or column-level policies, the
  enforcement should be amenable to providing partial data disclosure,
  which can also modify sensitive data before disclosure. 
  
\item \emph{Enable a clean separation between the policy
    specification, the application, and its underlying database:} 
  Given the scale of modern applications, enforcing policies as inline
  checks can quickly become difficult to maintain and implement
  correctly. Thus, it is imperative that the enforcement framework
  should be independent of the application and the database.

\item \emph{Practicality:}
  As a separate policy enforcement mechanism incurs overhead for
  policy selection and application, it should be a practical solution
  for the users offering very low or negligible performance overheads
  while correctly enforcing the policies. 
\end{enumerate}

In this work, we build on two key insights for enforcing contextual
and granular policies:

\noindent
1. \emph{Application-specific information needed to enforce
    fine-grained contextual policies can often be obtained at the
    API-endpoint boundaries of such applications:}   
  Typically, servers expose endpoints that a client can call into using
  a uniform resource identifier (URI). These endpoints map to
  different APIs defining distinct functionalities, which can further
  be mapped to what data is retrieved from the database and under what
  context the access occurs. As different APIs may 
  access the same data from a database under different contexts,
  policies that are \emph{API-specific} will reflect the necessary
  application context. For example, if a doctor in the emergency
  department of a hospital requests access to a patient's PHI through
  the emergency treatment endpoint (say, \textsf{/emergency/}), it can
  be reasonably inferred that the purpose of data access is an
  emergency. Thus, the context of data access can be determined based on
  the API endpoint used to access the data. Even if there isn't such a
  distinction of access, the application code can be refactored to
  introduce this distinction so that the application can correctly
  specify the policies.

  \noindent
2. \emph{Policies that selectively release information can be applied
    after query evaluation by mediating the results returned by the
    database:}
  While normal access policies can be applied before data access by
  query modification, policies that modify the content of the data
  cannot be enforced via query rewriting (when dealing with multiple
  policies - more in Section~\ref{sec:decpol}). Thus, a declarative
  policy specification in SQL is inadequate to enforce policies that
  release parts of a cell's content. However, such
  policies can be enforced by modifying the data returned by the
  database after query evaluation. 

We present {\framework}, a policy enforcement framework that addresses
the challenge of enforcing contextual and granular policies for
database-backed applications. In {\framework}, developers specify
policies that are explicitly associated with specific APIs provided by
their application. The policies are associated with the database schema
separately from the application code without requiring any support
from the back-end database. This makes it easier to implement, modify,
and audit both the policy and application.

{\framework} supports fine-grained contextual policies that are
factored into \emph{pre-eval} and \emph{post-eval}
components. Database accesses throughout the application are subject
to pre-eval policies, which are enforced before database access by
query rewriting. Post-eval policies provide the flexibility 
needed by partial release or declassification requirements, by
modifying the results of a database query at the granularity of
individual rows. 

We demonstrate that {\framework} can be applied to real 
applications to enforce a diverse range of practical data access 
policies. We prototyped {\framework} in Python, on top of Django, and
applied our prototype to migrate and build applications with complex
policy requirements: a port of the open-source Spirit forum
software~\cite{spirit} and open-source social-networking site
Vataxia~\cite{vataxia}, modified to  enforce fine-grained data flow
policies that restrict how users access topics, posts and
user-profiles; a conference management system built on top of
Jacqueline~\cite{yang2016} used for an academic workshop ported to
{\framework}; and as a microbenchmark, we build an intranet
application that manages the profile and compensation details of
employees, and facilitates events and meetings. Using these
applications, we show that {\framework} incurs very low overhead over
the original applications where policies are inlined through the
code.  

\subsubsection{Contributions}
To summarize our contributions, we present a novel policy enforcement
framework, {\framework}, that supports rich contextual policies on
sensitive data while maintaining strong separation between policy,
code and data. {\framework} factors policies into pre- and post-eval
components, enabling both data access and information release
restrictions in policies. {\framework} does not require
modifications to the application or the database, thus simplifying
policy implementation and maintenance. We prototype an enforcement
mechanism for {\framework} policies as an integrated language
runtime in Python on top of Django and evaluate it by using it to
build/migrate four applications ranging from 1000 LOC to 80 kLOC. We
show that it incurs low overheads while requiring minimal changes to
existing application code. 


\section{{\framework} Overview}
\label{sec:examples}
{\framework} is a framework that assists the developer in building
policy-compliant applications. In {\framework}, a policy is specified
centrally alongside the database schema, and the runtime ensures that
the policy is enforced correctly across all application components. 
{\framework} supports fine-grained, contextual policies that are
factored into \emph{pre-eval} and \emph{post-eval}
components. Policies are enforced based on which tables and fields
(and their transformations) are queried.

\subsection{Policy specification in {\framework}}
Policies in {\framework} have the following form:
$$\textsf{t}_1.\textsf{f}_1, \ldots,
\textsf{t}_n.\textsf{f}_n;~\varphi;~\api : \policy$$  
In this expression, $\policy$ is the policy body containing executable
code that either rewrites a query or filters a set of results; $\api$
is an optional API identifier; $\varphi$ is either \emph{pre} or
\emph{post}; and $\textsf{t}_i.\textsf{f}_i$ is a column identifier
from the target database schema. 

The policy $\policy$ applies when $\textsf{t}_1.\textsf{f}_1, \ldots,
\textsf{t}_n.\textsf{f}_n$ are accessed in a query. $\varphi$
indicates whether the policy is a pre-eval policy
(\emph{\textsf{pre}}) or post-eval policy (\emph{\textsf{post}})
depending on which the policy is applied before or after the query is
executed. 
The function $\policy$ works on either an
unevaluated query for pre-eval policies, i.e., before the data is
actually fetched from the database, or rows in the result-set of an
evaluated query for post-eval policies. $\policy$ always returns a
modified query or row, respectively. 

\paragraph{Pre-eval policies: }
Database accesses throughout the application are subject to pre-eval
policies, which are enforced before query evaluation by modifying
queries to have additional conditions. These policies add additional
filters to the query by either adding conditions to the query's
\textsf{WHERE} clause or inserting \emph{subqueries}, thereby limiting
the information returned by the database and the rows being accessed
by the query.   

\paragraph{Post-eval policies: }
{\framework} supports post-eval policies that provide the necessary
flexibility by associating policies with the result of an evaluated
query. Such policies operate at a finer-granularity with more
contextual information, and are used alongside generic pre-eval
policies for complete mediation. They are mainly information release
policies that apply on the query's result, and modify the rows in the 
result of the query or the result itself. Only the field(s) for which
the policy is specified is(are) modified while other fields in the
result remain unchanged.

\medskip
Policies can additionally be associated with a
transformation function on a field, e.g., \textsf{Avg},
in which case one of the $\textsf{t}_i.\textsf{f}_i$ is
replaced with $\textsf{F}_i(\textsf{t}_i.\textsf{f}_i)$ where
$\textsf{F}_i$ is the transformation used on
$\textsf{t}_i.\textsf{f}_i$. 

The field $\api$  contains the list of APIs on which the
current policy applies. Based on the API through which the data is
accessed, the policies with $\api$ take precedence and
override other policies. If $\api$ is omitted, then
the policy is applied for every access. 

The policies, additionally, have access to the current user
authenticated with the system and the API that made the query to the
database, which are extracted from the request sent to the server from
the client. The current user is represented as $\user$ in the
policies. If the user is not authenticated with the server, the user
is treated as an  anonymous user. The policy selection algorithm
checks if the current API is present in the $\api$ field of the policy
and returns the policies that apply for that API. 

In case policies are specified for $\textsf{t}_1.\textsf{f}_1$,~
$\textsf{t}_2.\textsf{f}_2$, and $\textsf{t}_1.\textsf{f}_1,
\textsf{t}_2.\textsf{f}_2$,  all three policies are applied when
accessing $\textsf{t}_1.\textsf{f}_1, \textsf{t}_2.\textsf{f}_2$. 
If no policy applies on a query, the default policy returns no rows to
account for missed policies.  

\subsection{{\framework} through Examples}
Using an intranet for a large organization as a running example, the
remaining section describes example {\framework} policies and their
specifications. We start by showing simple access control policies
followed by how more complex policies are enforced, as the 
section proceeds. The intranet provides different services to
employees like viewing personal employee details, payroll information,
and setting up meetings and events within the organization. The schema
for the back-end database of this application is shown in
Table~\ref{tab:schema}. Briefly, the \textsf{User} field contains the
personal details of the company's employees; the \textsf{Payroll}
field stores the details of employee salaries along with their
manager's identifier; the \textsf{EventCalendar} field records the
events or meetings organized within the company while \textsf{Invitee}
stores the list of employees invited to each event.  
\begin{table}[b]
  \centering
  \begin{tabular}{|l|l|}
    \hline
    \multicolumn{1}{|c|}{\textbf{Table}} & \multicolumn{1}{|c|}{\textbf{Fields}}  \\ \hline
    \textsf{User}                               & \textsf{id, name, age, address, dept}      \\
    \textsf{Payroll}                            & \textsf{id, mgid, salary}                  \\    
    \textsf{EventCalendar}                      & \textsf{eid, date, location, orgid, event} \\ 
    \textsf{Invitee}                            & \textsf{eid, empid} \\ \hline
  \end{tabular}
  \vspace{2mm}
\caption{Database schema for a company's intranet}
\label{tab:schema}
\end{table}


In the examples that follow, we use an object-oriented policy
specification that employs pre-defined methods (e.g., \textsf{filter}, 
\textsf{exclude}) to specify policies for enforcement on SQL
queries. Evaluation of a query executes it on the database and 
retrieves the relevant rows as the result. The pre-eval policies work
on a query (\textsf{query} in the policy) and are applied before the
query is evaluated. On the other hand, the post-eval policies apply on
the result of evaluation of a queryset (\textsf{result} in the
policy). Policies have access to the current user authenticated
with the server (denoted $\user$ in the policy). 


\medskip
\paragraph{\textbf{Example 1 (Access control policies): }}
Suppose that the site enforces a policy that allows either the
\textsf{name} or the \textsf{age} of the employees to be accessed
separately, but data linking the \textsf{name} and the \textsf{age}
should only be accessible to the employee whose \textsf{name} and
\textsf{age} is being accessed, or to an employee from the \textsf{HR}
department. This is a basic access-control policy that defines what
data is accessible by a user in terms of the database state, and can
be specified as a pre-eval policy in {\framework}. The annotation
\emph{\textsf{pre}} in the first line of the policy that follows the list
of columns on which the policy applies, identifies the policy as
pre-eval, as shown below: 
\begin{lstlisting}
User.name, User.age; (*\emph{pre}*) :
  if (*$\user$*).dept != 'HR':
     query = query.filter(id=(*$\user$*).id)
  return query
\end{lstlisting}
The policy above checks if the current user (identified by $\user$) is
in the \textsf{HR} department. If so, the policy doesn't add any
filters to the original query. If not, the policy ensures that the
user accesses only his/her own record. The variable \textsf{query}
refers to the query \emph{object} that will be executed on the
database while the function \textsf{filter} adds additional
constraints on the query. In the above example, the function adds a
clause to remove those rows whose \textsf{id} is not equal to the
current user's \textsf{id}.  

In certain cases, instead of denying access entirely, the application
needs to release some information about a sensitive datum such as an
aggregate statistic or a derived value. These policies can be
expressed as pre-eval policies in {\framework}. In our running
example, suppose that the policy requires that non-manager
employees can only access the average for other non-manager employees: 
\begin{lstlisting} 
Avg(Payroll.salary); (*\emph{pre}*) :
  mgr = Payroll.values('mgid')
  if (*$\user$*).id not in mgr:
     query = query.exclude(id__in=mgr)
  return query
\end{lstlisting}
The above policy is enforced when the average of employee salaries
(\textsf{Avg(Payroll.salary)}) is accessed by an employee. The policy,
initially, retrieves the employee-ids of all managers in the
organization. The function \textsf{values} returns only those fields
in the table (in this case, \textsf{Payroll}) that are specified as an
argument to the function, i.e., \textsf{mgid} (the manager's id). If
the current user is a manager, it returns the average of salaries of
all employees. Otherwise, it adds a clause to remove the salaries of
employees who are also managers in the query, before computing the
average, using the \textsf{exclude} function. 

\medskip
\paragraph{\textbf{Example 2 (Context-dependent partial release): }}
There are other cases where the developer might want to release
partial information about the sensitive information, which might vary
according to the context. In {\framework}, these policies are most
naturally expressed as post-eval policies that are applied to a set of
query results using the \textsf{result} object. This is denoted with
the \emph{\textsf{post}} annotation at the top of the policy.

Suppose the application enforces a policy that the address of an
employee is selectively visible to other employees, such that only the
employee can see his/her complete address, employees in the
\textsf{Transportation} department can see the neighborhood of an
employee to arrange a drop-off, and all other employees can only see
the city name. As pre-eval policies apply on the query and
not on individual rows, it is not possible to perform different
transformations on different rows. The policy given below demonstrates
how a post-eval policy can post-process results to enforce this finer 
constraint: 
\begin{lstlisting}[numbers=left] 
User.address; (*\emph{post}*) :
  for row in result:
    if row.id == (*$\user$*).id: (*\label{degex:if}*)
      continue
    elif (*$\user$*).dept == "Transportation": (*\label{degex:eif}*)
      row.address = getngh(row.address)
    else: (*\label{degex:e}*)
      row.address = getcity(row.address)
  return result
\end{lstlisting}
This example returns different values based on the context. If the
current user is the employee itself, then the user is allowed access
to the complete address (line~\ref{degex:if}). If the user is in the
\textsf{Transportation} department (line~\ref{degex:eif}), the user
can see the neighborhood of the employee's address. If none of the two
conditions hold, the user sees only the city of the employee. 

\medskip
\paragraph{\textbf{Example 3 (API-specific differential access): }}
A more interesting scenario arises when the developer wants to specify
a policy that indicates only the existence of a sensitive value in
certain cases, and in other contexts reveals more information. Such
policies are not supported by any of the existing mechanisms. An
example of this is a calendar application, wherein the app allows
users to create new events and invite other employees to the events.  

\begin{figure}
\begin{lstlisting}[numbers=left, caption=APIs exposed by event calendar service, label=ecapi]
def get_events(request):
  e = EventCalendar.all()
  ...    
def delete_events(request):
  e = EventCalendar.all()
  ...    
def get_location_events(request, loc):
  e = EventCalendar.filter(location=loc)
  ...
\end{lstlisting}

\begin{lstlisting}[numbers=left, caption=Policies enforced by event calendar service, label=ecpol] 
EventCalendar.*; (*\emph{pre}*) : (*\label{ecpol:ql}*)
  return query.filter(eid__in=Invitee.filter(empid=(*$\user$*).id).values('eid'))

EventCalendar.*; (*\emph{pre}*); [delete_events] : (*\label{ecpol:qldel}*)
  return query.filter(orgid=(*$\user$*).id)

EventCalendar.*; (*\emph{pre}*); [get_location_events] : (*\label{ecpol:defpre}*)
  return query

EventCalendar.*; (*\emph{post}*); [get_location_events] : (*\label{ecpol:policyexc}*)
  for row in result:
    if not Invitee.filter(eid=row.eid, empid=(*$\user$*).id).exists(): (*\label{ecpol:inv}*)
      row.event = "Private event" (*\label{ecpol:false}*)
      row.orgid = 0
  return result
\end{lstlisting}
  \vspace*{-5mm}
\end{figure}

Consider three endpoints identified by APIs \textsf{get\_\-ev\-ents},
\textsf{delete\_\-events} and \textsf{get\_\-loc\-ation\_\-events}
shown in Listing~\ref{ecapi}. The API \textsf{get\_\-ev\-ents} returns
a list of events, the API \textsf{delete\_\-events} returns a list of
events that can be deleted, and the API
\textsf{get\_\-loc\-ation\_\-events} returns the list of events
organized at the location identified by \textsf{loc}. The function
\textsf{all} retrieves all rows from the table. The system
enforces a pre-eval policy that allows a user to see only those events 
that the user is invited to.  When accessed through the
\textsf{delete\_\-events} API, the system enforces a pre-eval policy
that allows the user to see only those events that the user
created. However, when viewing the list of events at a location, the
user besides getting the details of the events to which she is invited
should also see \textsf{``Private event''} for other events along with
their date and time so that she cannot schedule another event at the
same location at the given time. 

As getting a list of events is subject to the pre-eval policy that
returns only those events that the user is invited to, a policy needs
to be specified specifically for APIs accessing events to delete some
events, e.g., \textsf{delete\_events}, and at a location, e.g.,
\textsf{get\_\-loc\-ation\_\-events}. The list of specified policies
is shown in Listing~\ref{ecpol}. For API-specific policies, the list
of APIs that these policies apply to are specified after the
\emph{\textsf{pre}} or \emph{\textsf{post}} annotation. If the value
is omitted, the policy applies to all APIs. 

When the list of events is accessed via \textsf{get\_events}, it
applies the pre-eval policy defined on line~\ref{ecpol:ql} and returns
only those events that the user is invited to. The `$\ast$' in the
policy indicates that the policy applies every time the table is
queried. When the list of events is accessed via
\textsf{delete\_\-events}, it applies the pre-eval policy defined on
line~\ref{ecpol:qldel} overriding the policy defined on
line~\ref{ecpol:ql} as it is API-specific, and returns only those
events that the user has created. When the list of events at a given
location is accessed via \textsf{get\_\-loc\-ation\_\-events}, it
retrieves \emph{all} events from the database at that location as the
post-eval policy defined on line~\ref{ecpol:policyexc} overrides the
pre-eval policy. It checks if there is an entry in the
\textsf{Invitee} table that corresponds to the current row's event and
the current user (line~\ref{ecpol:inv}). If found, the policy returns
the details of the event to the user; if not, it returns
\textsf{``Private event''} for the event (line~\ref{ecpol:false}) with
the organizer's data hidden. The policy on line~\ref{ecpol:defpre}
ensures that the query is not filtered, thus, returning all events
registered at the given location. Without this, the policy on
line~\ref{ecpol:ql} would filter the list to contain only those events
that the user is invited to. 

\section{Policy Framework}
\label{sec:framework}
\begin{figure}
  \centering
  \includegraphics[width=0.3\textwidth]{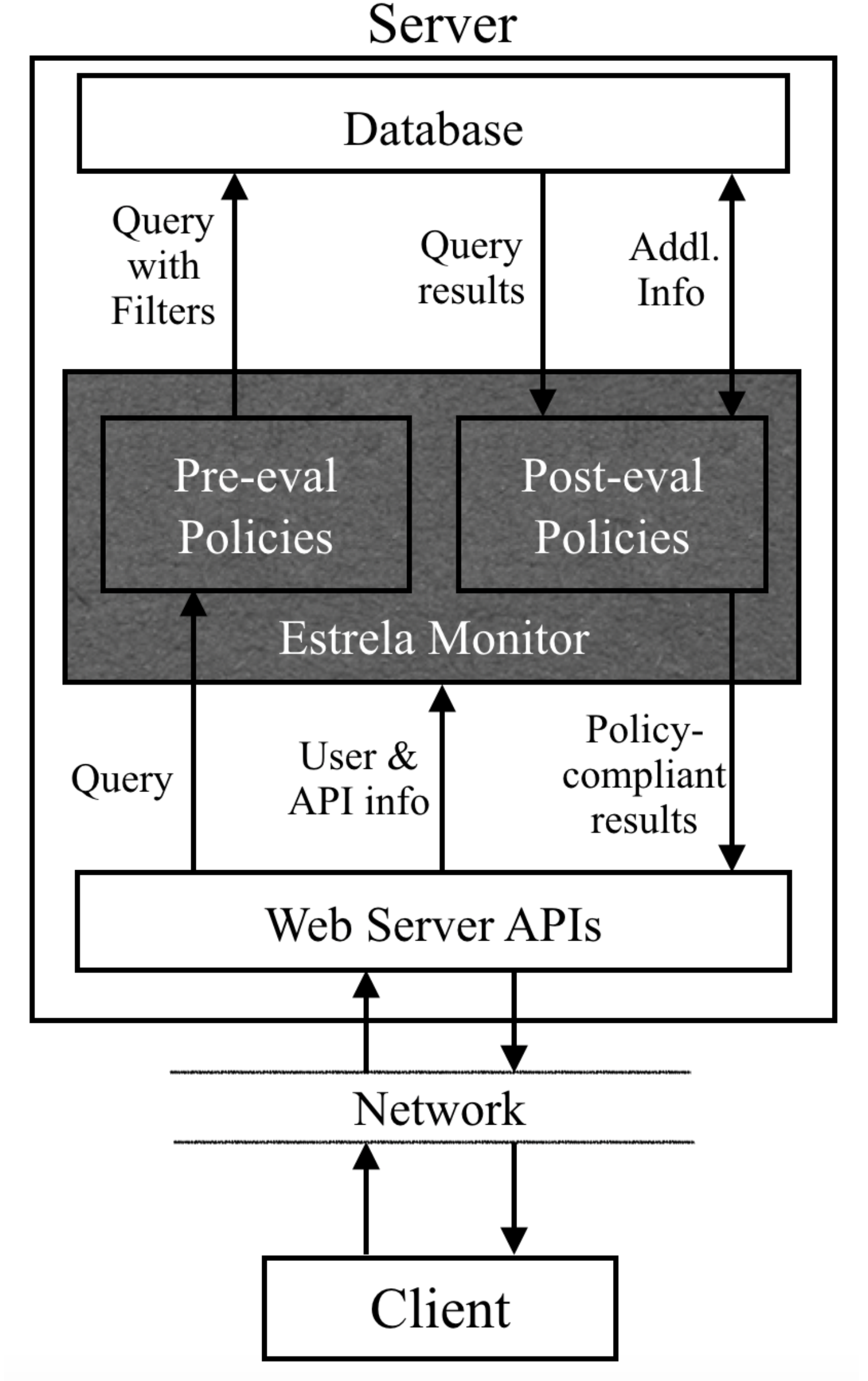}
  \caption{Application architecture with {\framework}}
  \label{fig:framework}
\end{figure}
\subsection{Application architecture with {\framework}}
Figure~\ref{fig:framework} describes the application architecture with
the policy enforcement mechanism using {\framework}. The overall
workflow of the architecture is as follows: 
\begin{enumerate}
\item The server accepts incoming requests on different APIs and
  performs the required query on the database. 
\item Pre-eval policies associated with the data being retrieved are
  applied as filters on the query before the object is fetched from
  the database.  
\item The filtered query is then executed on the database, which gets
  some data from the database. 
\item Once the data is fetched, the post-eval policies that apply on
  the result of the query modify the query's result-set.  
\item The API performs any necessary computations on the data
  retrieved, and creates and sends the response to the client. 
\end{enumerate}

\subsection{Threat model}
{\framework} aims to provide a unified, practical, and robust
framework for specifying and enforcing policies in database-backed
applications.  The primary security goal of {\framework} is to ensure
that the database queries and the APIs satisfy the conditions
specified in the policy. {\framework} mitigates authorization bugs by
enforcing data flow policies that apply when the data is read from the
database, and limit the scope of our paper to those. We assume that
developers make a good-faith effort to specify correct policies, and
that the integrity of the {\framework} framework is intact and
uncompromised throughout the lifetime of the application. 
{\framework} does not attempt to prevent leaks from hardware or
operating system-level side channels, or by groups of users who
collude via out-of-band channels to learn more than what is specified
in the policy. Likewise, as {\framework} policies concern server-side
data, leaks that result from vulnerabilities in the client-side
browser or operating system are also not in scope for our security 
goal. 

\subsection{Algorithm for policy enforcement }
\begin{algorithm}[!t]
  \begin{algorithmic}
    \Function{GetPolicy}{$P$, \textit{fields}, $A$, \textsf{def}}
    \State \textit{policy} = \{\}, \textit{epol} = \{\}
    \For{$(p:\varphi)$ \textbf{in} $P$}
    \If{(\textit{fields}; $A$) \textbf{in} $p$}
    \State \textit{policy} = \textit{policy} $\cup~\varphi$
    \ElsIf{(\textit{fields}) \textbf{in} $p$}
    \State \textit{epol} = \textit{epol} $\cup~\varphi$
    \EndIf
    \EndFor
    \If {\textit{policy} \textbf{is empty}}
    \If {\textit{epol} \textbf{is not empty}}
    \State \textit{policy} = \textit{epol}
    \ElsIf{\textsf{def}}
    \State \textit{policy} = $\varphi_{\mathbb{D}}$
    \EndIf
    \EndIf
    \State \Return{\textit{policy}}
    \EndFunction
    
    \Function{Apply}{$Q, A$, \textit{Pre, Post}}
    \State \textit{fields} = parse($Q$)
    \State \textit{policy} = \Call{GetPolicy}{\textit{Pre}, \textit{fields},
      $A$, \textsf{true}}
    \State $Q$ = applyFilters($Q$, \textit{policy})
    \State \textit{result} = executeQuery($Q$)
    \State \textit{postpol} = \Call{GetPolicy}{\textit{Post}, \textit{fields},
      $A$, \textsf{false}}
    \State \textit{result} = applyPostPolicy(\textit{result}, \textit{postpol})
    \State \Return{\textit{result}}
    \EndFunction
  \end{algorithmic}
  \caption{Algorithm to enforce policies in {\framework}}
  \label{alg:policy}
\end{algorithm}
Algorithm~\ref{alg:policy} shows the top-level algorithm (and the
auxiliary function) used by {\framework} to select and enforce
policies on data access. The top-level function \textsc{Apply} takes
the query, API, and the set of policies as the input. The function
starts by parsing the query ($Q$) to determine which fields are used
by the query. Using the \textit{fields} and the API information $A$,
the function \textsc{GetPolicy} returns the pre-eval policies
applicable to the query. These policies are enforced on the query,
which is then run. The results returned by the database are modified
as per the post-eval policies selected by \textsc{GetPolicy} for the
\textit{fields} and the API $A$. The modified result is then returned
to the user.

The operation (\textit{fields}; $A$) \textbf{in} $p$ in
\textsc{GetPolicy} checks if the fields and APIs in the policy $p$
contain \textit{fields} and $A$, i.e., the list of fields and APIs in
$p$ is a superset of \textit{fields} and $A$. If \textit{fields}
contain aggregate functions, it additionally checks for the fields
used in the aggregate function in the fields of $p$. If the
appropriate policies for the API $A$ are not found, the function
applies general policies that apply to all APIs. The policy
$\varphi_{\mathbb{D}}$ is the default policy that restricts access to
all the data. The variable \textsf{def} is set to \textsf{true} or
\textsf{false} to indicate whether the default policy should apply.

\subsection{Declarative vs. object-oriented policies}
\label{sec:decpol}
{\framework} enforces multiple policies associated with the (set of)
fields specified in the query. It filters data based on the pre-eval
policies \`a la the existing access-control mechanisms, and modifies
the data in the result to release partial information as per the
post-eval policies. In our prototype, instead of specifying policies
in a declarative language like SQL, we specify them using
object-relational mapping (ORM) methods that offer an object-oriented
view and a method-based access of the data. Besides 
making it easy to write policy functions for developers who might be
more comfortable using object-oriented programming for
policy-specification, ORM-based specification has certain advantages
over the declarative specification when specifying complex policies: 
\begin{enumerate}
\item Post-eval policies cannot be specified in SQL as they apply on
  a query's result requiring an imperative specification. 
\item Enforcing multiple policies when specified in SQL forces the
  policies to be access-control (as they are specified as additional
  conditions in the WHERE clause~\cite{mehta2017,LeFevre2004,Byun2008}
  or specific SQL sub-queries~\cite{LeFevre2004, Wang2007}); thus, not
  allowing information release unless using user-defined functions
  that are already defined in the database, which, in turn, requires
  modification to the application code. 
\item Enforcement of multiple data usage policies for information
  release (e.g., DataLawyer~\cite{upadhyaya2015}) require \emph{all}
  policies to be checked before the policy-compliant data is returned,
  making the approach less efficient.  
\end{enumerate}

\subsection{Discussion}
\subsubsection{Comparison with existing access control enforcements}
While the access control mechanisms enforced by enterprise database
systems~\cite{oraclevpd, ibmdb2, sepostgresql} provide fine-grained
data access, they have certain shortcomings that make them unsuitable
for applying the policies supported by {\framework}:
\begin{itemize}
\item They lack support for policies that link two or more tables
  without having explicitly defined views. Creating such views require
  modifications to the application such that it queries for the
  correct data.  
\item Data masking either removes all rows or defaults the values in
  the column if the rows are inaccessible by the user. They do not
  support modification of specific rows in the table as shown in
  Example 2 above.
\item Contextual policies are application-specific, and hence cannot
  be implemented generically by database access control
  systems. Existing systems~\cite{Byun2008, Kabir2011} that take into
  account such contexts when allowing access require explicit
  specification by the user in the query, which they validate. While
  the user needs to be trusted to provide the correct context, it also
  requires modifications to applications to send appropriate queries.
\end{itemize}

As most of the other existing approaches~\cite{mehta2017, upadhyaya2015,
  LeFevre2004, Wang2007, Rizvi2004, Byun2008, Stonebraker1974} work
independent of the application, they do not have access to the
API-specific information necessary for enforcing contextual
policies. It is, however, possible to expose the path or API on which
the request is made (similar to how the authenticated user is exposed)
to the enforcement monitor. This would, in turn, require the existing
approaches to modify the policy specification language to take into
account API information for enforcing contextual policies.
However, it is not possible to specify post-eval policies in the
existing approaches (making modification to query results difficult),
which is required, for instance, in Example 3 when fetching events at
a particular location (\textsf{get\_location\_events}).  

It is also instructive to see how the policy in Example 2 differs from 
prior work like Qapla~\cite{mehta2017}, which also requires
user-defined functions like \textsf{getngh} and \textsf{getcity} for
specifying such policies. In Qapla, the developer has to invoke the
correct function, e.g., \textsf{getngh(address)} if the user is in the 
\textsf{Transportation} department and \textsf{getcity(address)} if
the user does not satisfy any criterion, to get access to the
address. If the developer queries the wrong function, she does not get
access to the address.  In {\framework}, the post-eval policy takes
care of this; thus no modification to the application or its API is
required. Moreover, the query needs to be repeated to get the actual
values of address and the city name when the user does not have
appropriate access.

Other prior works~\cite{LeFevre2004, Wang2007} can use \textsf{CASE}
statements to specify such properties because their rewriting
technique generates different data based on the conditions. However,
when multiple policies apply on a query for a column, it is unclear as
to how the policies would apply in these approaches.

\subsubsection{Pre-eval vs. Post-eval policies}
Post-eval policies are at least as expressive as pre-eval policies,
but enforcing pre-eval policies has certain advantages. Firstly, as
pre-eval policies are applied before the query is evaluated by adding
filter conditions or subqueries, the database is queried only
once. This enhances the performance of the policy-framework as
post-eval policies require more database hits (as the policy queries
the database again). Secondly, pre-eval policies prevent
timing-related leaks that are possible with post-eval policies. As
they are applied before the database is queried, they do not reveal
any information about the number of records satisfying the original
unfiltered query. For instance, suppose, in a healthcare setting, a
user wants to know how many patients have a certain disease, but say,
they are not allowed to access this information. With post-eval
policies, the time taken to first retrieve the list of patients that
have a certain disease and then filtering the results might be
significant, dominating the time to respond to the query. Therefore,
if it takes a long time to respond, this may leak some information on
what the query response size was to the user, even though the result
itself contains no sensitive information. 

Post-eval policies allow partial release of sensitive information
based on the context making them flexible enough to handle cases where
different values need to be returned as per the context. Moreover, it
is not possible to enforce all policies before evaluation as some
policies need to post-process the results. Post-eval policies are also
useful when some additional information that is not present in the
result of the filtered query needs to be released.

\subsubsection{Preventing implicit leaks}
The policies are applied considering \emph{all} the fields used in the
query irrespective of where they appear in the query. While this may
at first seem too conservative, it is necessary to prevent implicit
information leaks, as might be the case when, for example 
a query returns the names of all employees of a particular age
(\textsf{SELECT name FROM employees WHERE age > 45}). The policies for
accessing \textsf{name}, and \textsf{name} and \textsf{age} together
may be different --- while all employees can access the names of other
employees, an employee can only access his/her own age and name
together as shown below:
\begin{lstlisting}[numbers=none]
User.name; (*\emph{pre}*) :
  return query
User.name, User.age; (*\emph{pre}*) :
  return query.filter(id = (*$\user$*).id)
\end{lstlisting}
If the policy were based only on the selected \textsf{name} column, it
would apply a relaxed policy that allows the names of all employees
having the particular age to be displayed as the first policy does not
apply any filter for access of \textsf{name}. However, as the two
columns are linked in the query, the correct policy would be the
second one associated with both \textsf{name} and \textsf{age} that
reveals to the user his/her own name and age only and no additional
information because of accessing name and age together. To handle such
leaks, all columns used in the query are considered when selecting the
policy to apply on the query.

\section{Implementation}
\label{sec:enforcement}

{\framework} provides a mechanism for automatically enforcing a
given set of policies. We discuss its implementation in this section
and evaluate the approach in Section~\ref{sec:eval}. 

{\framework} is prototyped in Python and extends Django~\cite{django},
a Python-based model-view-template application framework. 
Thus, apps written in {\framework} are otherwise standard Django apps
with policies in the schema. Policies are specified alongside the
database schema using two class-methods (for pre-eval and post-eval
policies) that are inherited by all models (schemas). If these methods
are not overriden by a model, then a default conservative policy that
suppresses all results applies. It includes an object-relational
mapping (ORM) to interact with databases based on which it constructs
SQL queries, represented as \emph{querysets}. Evaluation of a queryset
object executes the query on the database and retrieves the relevant
rows as the result. Django constructs a queryset as soon as an API
starts querying for some data in the database; however, it does not
evaluate the queryset to reduce the number of database hits.
The pre-eval policies work on a queryset (\textsf{query} in
the policy) while the post-eval policies apply on the result. Policies
are Python functions in our prototype.

To enforce a pre-eval policy, we augment the database interface
functions provided by Django via monkey-patching, which amounts to
inheriting from Django's classes and overloading the methods relevant
to Django's interfacing with the database. We modified about 700 lines
of Python code including code from the original implementation.
Using this approach, we achieve complete mediation such that all
database accesses in an {\framework} application 
occur through the instrumented methods. These methods are responsible
for invoking the policy functions, and passing them the current user
and API information that {\framework} exposes from the request to the
server. As Django supports lazy evaluation, we apply the pre-eval
policies just before the query is evaluated to get a better
performance.   

To enforce a post-eval policy, we modify the results returned from the
database and selectively apply the relevant policy functions before
the results are returned to the APIs. Selective enforcement is
achieved by a case analysis on the fields involved in the query, to
determine which policies are relevant. When multiple policies apply on
the result of an API call, we assume an arbitrary but fixed order in
which to apply the policy functions to the queryset result. 

Both pre- and post-eval policies need to consult the set of active
API-specific policies to bypass enforcement when operating in the
context of a relevant API. 

\subsection{Integration with legacy applications}
As {\framework} is built on top of Django, migrating an existing
Django application to {\framework} is straightforward and does not
require changes to the core application code. The developer only needs
to specify the policies alongside the data models (i.e., the database
schema), which are modified to inherit from {\framework}'s model
class. The only other necessary change is to expose the request
parameters of each API to the policy enforcement mechanism, which is
taken care of automatically by {\framework} using a middleware
configured in the application's settings. 

Although {\framework} is prototyped in Python on top of Django, its
principles are generic and can be extended to enforce authorization
policies in other existing frameworks (designed using any
language). {\framework} requires a separation of the models or the
database schema from the actual application alongside which the
policies are specified in the language in which the framework is
built. Enforcing policies would require adding hooks in the query
evaluation process of the framework. 

An interesting question to consider when integrating {\framework} with
legacy applications is how to bootstrap policy specification to assist
the developers. Policy inference, which is orthogonal to the problem
studied in this paper, has been an area of active
research where prior works have proposed approaches to  mine
meaningful policies using logs, traces and program
specifications~\cite{han2000, bauer2011, xu12, xu2013, xu2014, xu2015, 
 is2015, bui2017, sacmat18, cotrini2018, sacmat19, acsac2019}. The
mined policies can be used to bootstrap the initial set of policies
when migrating applications to {\framework}. However, the problem of
policy inference is out of scope of the current paper.

\section{Evaluation}
\label{sec:eval}
We evaluate {\framework} by comparing the code changes required on
existing applications and the overhead it incurs due to the policy
checking. We demonstrate that {\framework} is easy to integrate with
existing applications, and incurs very low overheads, showing its
effectiveness and usefulness. We consider open-source applications for
migration to {\framework} that are built using Django.  

\subsection{Methodology and setup}
We used {\framework} to migrate a few applications to enforce
policies, ranging from about 1000 LOC to 80 kLOC. The first is a 
version of Spirit~\cite{spirit}, a forum software where 
users can discuss on different topics, migrated to {\framework} with
policy enforcement. The second application is a social-networking site
Vataxia~\cite{vataxia}. The third is a multi-user conference management 
system that lets users add, edit, remove papers for a conference. The
fourth application is a company's intranet on the lines of the
examples discussed in Sec.~\ref{sec:examples}.

The case-studies were chosen to evaluate the effect of policy
enforcement on large applications, applications with multiple
policies, and the overhead it incurs on simple and complex
policies. For all the case-studies, we add a middleware to expose the
request details  to the schema for evaluating the policies in
{\framework}. All other code in the applications remains the same. The
baseline implementations of our case-studies are all built using
Django with policies included in the code, allowing us to evaluate the
performance of  {\framework}. 

We performed our experiments on a MacBook Pro having an 8 GB RAM and 
3.1 GHz Intel Core i5 processor, running macOS Catalina 10.15.1. We
automated the process of sending requests to the server to retrieve
data from the database, and measured the average time taken for
processing the request and policy enforcement over 100 trials. The
back-end database was MySQL version 8.0.18. Unless
mentioned otherwise, all servers were build using Python 3.7.3
and Django 2.2.7. The server and the client run on the same machine so
the evaluation results do not include network latency or the time
taken for rendering. We ran this process in the context of 
a particular user, who is authenticated before sending the request.
The error bars in the graphs show the standard deviation. 

In the following sub-sections, we describe the functionality of these
applications, the policies that we apply, and how these policies are
implemented and enforced using {\framework}. We report the performance
numbers for these case-studies, by measuring the time taken by the
application to perform user operations in different
scenarios. Section~\ref{sec:code-sf} discuss the code changes required
for the implementations to enforce policies in their respective settings. 

\subsection{Case study: Spirit forum software}
Spirit is a Django-based forum software for facilitating conversations
and discussions amongst users. The original software contains about 80
kLOC. Users can create new conversations or post comments on existing
conversations depending on the visibility of the conversation and/or
whether the user has been invited to the conversation or not. The
schema of the actual site contains 28 tables with 165 columns. We
specified pre-eval policies for various tables and measured the
overhead incurred by {\framework}, and demonstrate the ability of
{\framework} to scale to large applications.  

\subsubsection{Migration effort}
\label{sec:code-sf}
We modified the models' base class in the original software to use
{\framework}'s model as the base class. This required a modification
of about 30 LOC in the original software. The other modification
required apart from the specification of policies was to expose the
incoming request to the models for identifying the current user and
the API that requested the data, which require a couple of lines of
code to be added to the settings of the application.    

One of the policy we enforce is associated with various topics in a
forum that allows only logged-in users to view the topics. Without the
policy enforcement, Spirit shows all public topics even without a user
being logged-in to the system. In {\framework}, we added a policy that
checks if the current user is authenticated or not; if not, it does
not show any topics to the user. In the original version of Spirit,
this check has to be propagated to at least three different files in
the codebase, all of which access a topic and display it to the user
reiterating the need for centralized specification of policies. 

\subsubsection{Performance}
\begin{figure}[]
  \centering
    \includegraphics[width=0.75\linewidth]{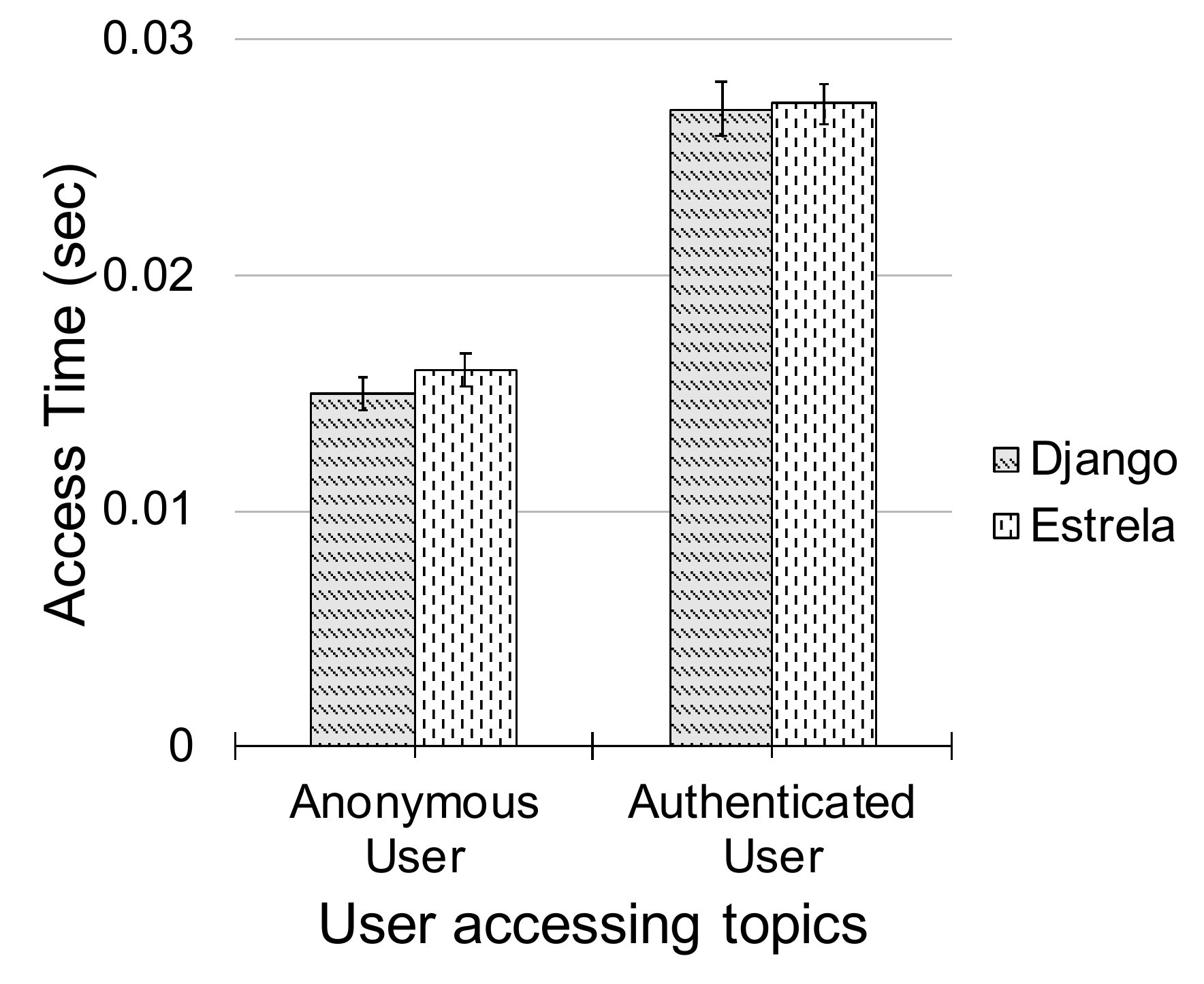}
  \caption{Time taken to request and access different topics in the forum with Django as the baseline}
  \label{fig:forum}
\end{figure}
We added 1000 users and 1000 topics to the database and evaluate the
time taken to access the topics with different users. The result of our
experiment is  shown in Figure~\ref{fig:forum}. {\framework} incurs an
average overhead of about $0.2$ ms or $0.8\%$ when accessing the
topics as an authenticated user. When accessing the topics as an
anonymous user, {\framework} incurs an overhead of about $1$ ms, 
which is mainly due to Django's lazy evaluation in
{\framework}. With {\framework}, the policies are applied just before
the evaluation after the objects have be created even if the objects
are never required in the future resulting in the additional
overhead. The inline check in Django can be added as early as possible
in the API code preventing the creation of objects, thereby avoiding
the additional overhead.

\subsection{Case study: Social-networking site}
\label{sec:sub:social}
Social-networking sites involve multiple users posting information and
sharing it with other users. Users can track the activity of other
users by ``following'' them. We modify an existing open-source
social-networking site, Vataxia~\cite{vataxia}, by implementing
additional functionality to follow users and view a user's posts. The
application allows new users to be added to the system and to search
for users to follow their posts. The original schema contained the
following models -- \textsf{User}, \textsf{Post},
\textsf{PrivateMessage}, \textsf{Reply} and \textsf{Vote}; we extend
\textsf{User} by adding a field, \textsf{follow}, to include a list of
users that the user is following. The table \textsf{Post} contains a
field \textsf{user}  referencing \textsf{User} table, and the message
that the user has posted in \textsf{msg}. The front-end of the
application is written in ReactJS while the back-end is developed
using Django REST framework~\cite{drf}, Django 2.2.7 and Python
3.7.3. 

\begin{figure}[h]
\begin{lstlisting}[caption=Application code for returning all posts and profile of a user, label=egSoc, basicstyle=\sffamily]
def posts_view(request):
  posts = Post.objects.all()            
  return Response(Serialize(posts))

def profile_view(request, uid):
  profile = User.objects.filter(id=uid)            
  user_posts = Post.objects.filter(user=uid)
  return Response(Serialize(profile, user_posts))
\end{lstlisting}
\begin{lstlisting}[caption=Policy specified in Django for Vataxia, label=egSocCode, basicstyle=\sffamily]
def posts_view(request):
  u = request.user
  if not u.is_authenticated():
    posts = []
  else:
    f_ids = u.follow.values('id')
    posts = Post.objects.filter(user__in=f_ids) (*\label{eg:soc1}*)
  return Response(Serialize(posts))

def profile_view(request, uid):
  u = request.user
  if not u.is_authenticated():
    profile = None
    user_posts = []
  else:
    profile = User.objects.filter(id=uid) 
    f_ids = u.follow.values('id')
    user_posts = Post.objects.filter(user=uid, user__in=f_ids) (*\label{eg:soc2}*)
    if not user_posts:
      user_posts = ['Follow user to see the posts']
  return Response(Serialize(profile, user_posts))
\end{lstlisting}
\begin{lstlisting}[caption=Policy enforcement for Vataxia in Estrela, label=egSocEst, basicstyle=\sffamily]
Post.*; (*\emph{pre}*):  
  if not (*$\user$*).is_authenticated():
    query = query.none()
  else:
    f_ids = (*$\user$*).follow.values('id')
    query = query.filter(user__in=f_ids)
  return query

Post.*; (*\emph{post}*); [profile_view]:    
  if not result:
    result = ['Follow user to see the posts']
  return result
\end{lstlisting}
\end{figure}

\subsubsection{Migration effort}
\label{sec:code}
We port Vataxia to {\framework} by adding policies for various tables
modifying the models of the application to inherit from {\framework}'s
base class. The original application contains around 19 kLOC as part
of the front-end and 2 kLOC in the back-end. We modify 10 lines in the
application, and include the middleware that exposes requests to the
models.

We discuss the enforcement of a policy that limits the posts that
a user can view based on whether the user is following the user whose
posts are being accessed or not. If accessing only the posts, a user
can see posts by only those users that she is following. However, when
the user accesses a profile of a particular user, she gets a default
message asking her to follow the user to see the posts. In
{\framework}, this policy is implemented as a post-eval policy
specific to the \textsf{profile\_view} API because when accessed
through the \textsf{posts\_view} API, this post-eval policy should not
be enforced. 

For code comparison, the policy codes for Vataxia are shown in  
Listings~\ref{egSocCode} and~\ref{egSocEst} for policy
in views, and schema with Estrela, respectively. 
Although the policies are similar across the two Listings, the
developer has to add the policy for every access of \textsf{Post} when
adding checks in the API code (e.g., on line~\ref{eg:soc1} and
line~\ref{eg:soc2} in Listing~\ref{egSocCode}) while the policy
specification is much cleaner and does not require any modification to
the code in Listing~\ref{egSoc}. 

\subsubsection{Performance}
\begin{figure}[]
  \centering
    \includegraphics[width=\linewidth]{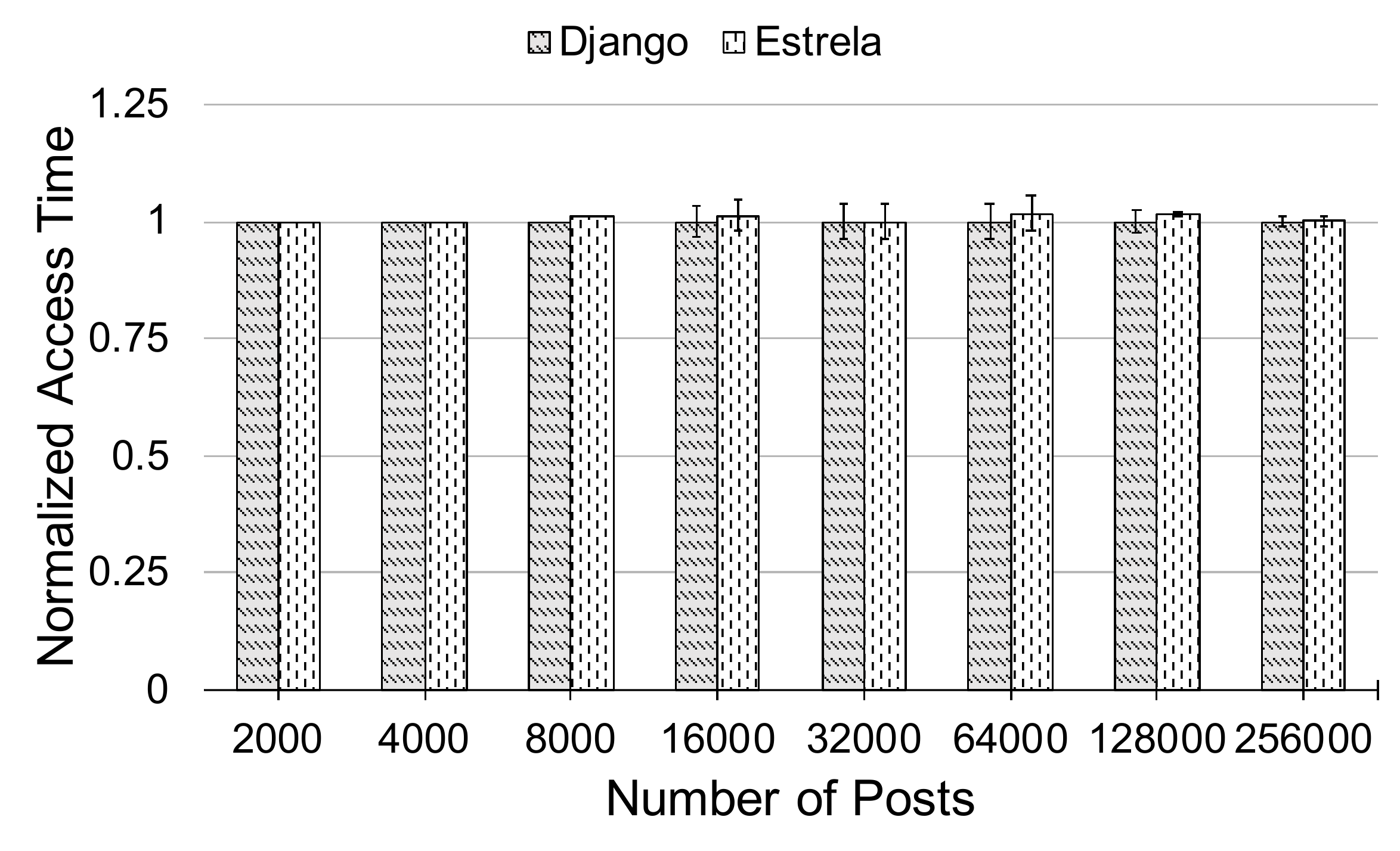}
  \caption{Normalized time taken to access all posts through a REST API with policy in code as the baseline}
  \label{fig:eval-all}
\end{figure}
We added $1000$ users and their details to the database for the
social-networking site with every user following $0-3$ other users in
the system. We measure the time taken to access posts in the system
for different number of posts (ranging from $2,000$ posts to $256,000$
posts). The posts are generated automatically before the processing
starts and are related to random users. Figure~\ref{fig:eval-all}
shows the normalized time taken in different scenarios. The average
overhead for accessing all posts with {\framework} when compared to
the baseline of policy in the application code is around $0.7\%$. 
The policy in {\framework} applies on the result before it is
returned, but incurs the overhead due to the additional check for
authentication of the user and selecting policies; the
enforcement time for the policies is almost the same as with policies
intertwined with the code. 

\begin{table}[h]
  \centering
  \begin{tabularx}{\linewidth}{|c|X|}
    \hline
    \multicolumn{1}{|c|}{~~~~~~\textbf{Table}~~~~~~} & \multicolumn{1}{c|}{\textbf{Policy - User can access}} \\ \hline
    \textsf{User} & \textsf{either her own profile or another user's profile if she is the chair or in the pc}      \\
    \textsf{Paper} & \textsf{any paper if she is the chair}                  \\    
    \textsf{Paper} & \textsf{non-conflicted papers if she is in the pc}                             \\
    \textsf{Paper} & \textsf{papers which she has co-authored} \\ 
    \textsf{Paper} & \textsf{accepted status of a paper either if she is the chair, or if she is the co-author and phase is final} \\ 
    \textsf{Paper} & \textsf{number of submitted and accepted papers in final phase} \\
    \hline
  \end{tabularx}
  \vspace{2mm}
  \caption{Policies for the conference management system}
  \label{tab:confpol}
\end{table}

\subsection{Case study: Conf. management system}
We modify an existing conference management system 
built in Jacqueline~\cite{yang2016}, using Django, that has been
deployed to manage paper reviewing process for an academic workshop
(PLOOC 2014), to work without information flow tracking in place and
migrate it to {\framework}. This modified application contains about 4
kLOC retaining the features of the existing system like creating users
and conferences, adding papers and roles for users, etc. while
removing its dependence on the Jeeves~\cite{yang2012} library and the
functionality to handle faceted values; the database also does not
contain additional fields for handling facets. The system supports
single-blind submission, handling conflicts, and submission of reviews
and comments by reviewers. Every conference has three phases -
submission, review and final -  which influence the policies. We
enforce the policies shown in Table~\ref{tab:confpol} 
on different tables in the schema. 

\subsubsection{Performance}
\begin{figure}
  \centering
    \includegraphics[width=0.5\linewidth]{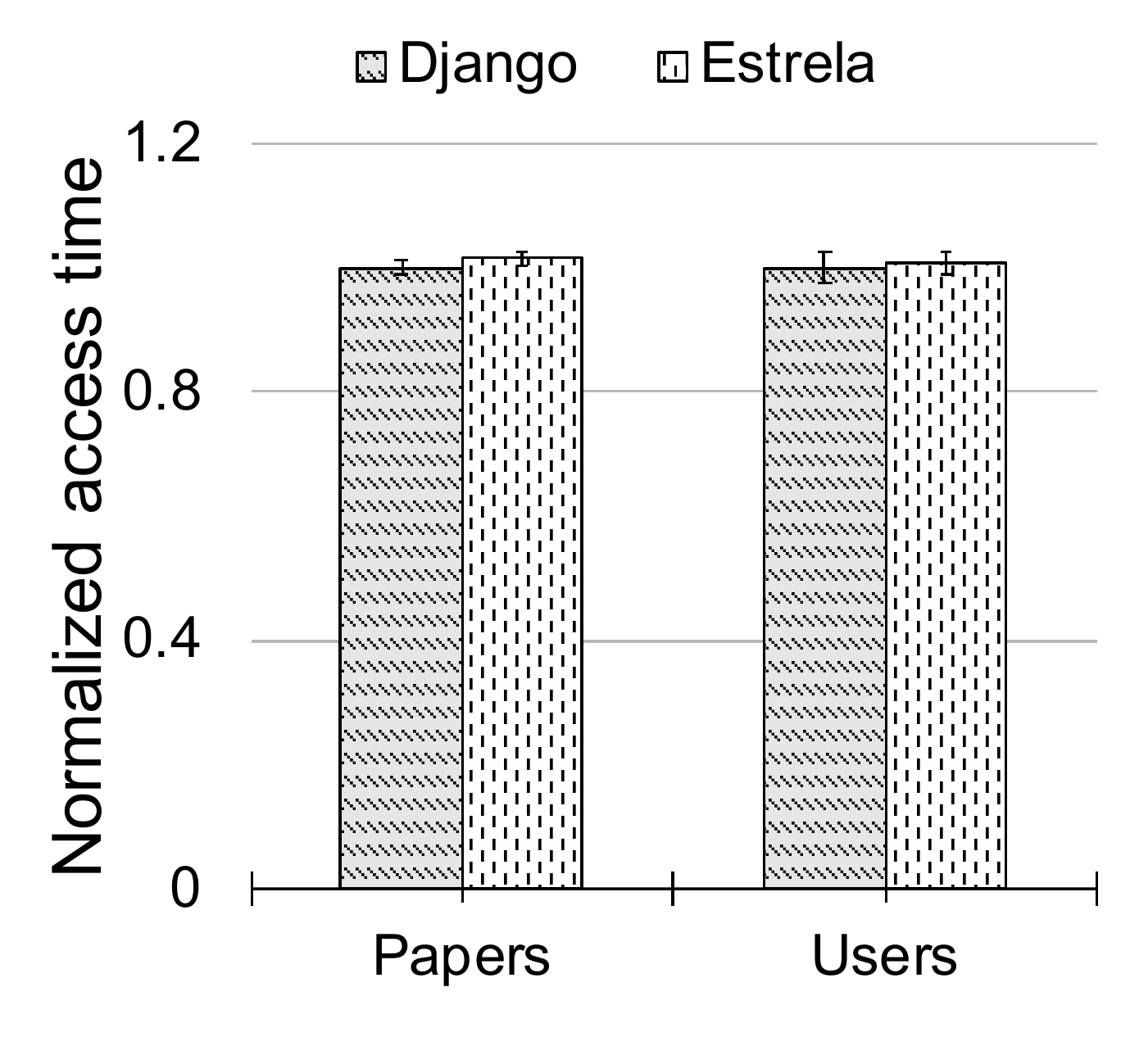}
  \caption{Normalized time taken to access all papers and users with Django as the baseline}
  \label{fig:eval-conf}
\end{figure}
We created a dummy conference with $1000$ users (25 of them on the
pc), and 1000 papers submitted by randomly chosen authors with 2
co-authors each. The policy for \textsf{User} table shown in
Table~\ref{tab:confpol} is a pre-eval policy while \textsf{Paper} has
a mix of both. Figure~\ref{fig:eval-conf} shows the normalized time
taken to access all papers and users in the system with Django as the
baseline. The performance of {\framework} when accessing the list of
users incurs an overhead of about $0.6\%$. {\framework} incurs an
additional overhead of around $1.5\%$ as compared to Django when
accessing the list of papers due to the number of policies (both
pre-eval and post-eval) associated with the \textsf{Paper} table.  

\subsection{Case study: Company intranet}
As a microbenchmark to evaluate the performance of {\framework} when
enforcing different kinds of policies shown as examples in
Section~\ref{sec:examples}, we implement an intranet website in
{\framework} that handles employees' personal and official
information. The employees of the company can access their profiles,
payroll information, their friends' profiles, and can schedule events
or meetings within the company as described earlier in
Section~\ref{sec:examples} extended with an additional
\textsf{Friends} table that contains a list of friends in the
organization. A brief description of the policies that we enforce is
shown in Table~\ref{tab:queries}. The policy-compliant queries
executed on the database through Django and {\framework} are the
same. As policies in {\framework} use high-level functions that are
also used for the inline checks in Django, the modified query
generated in {\framework} is the same as the one generated through
Django using inline checks. 

\begin{table}
  \centering
  \begin{tabularx}{\linewidth}{|c|X|}
    \hline
    \multicolumn{1}{|c|}{~~~\textbf{Query}~~~} & \multicolumn{1}{c|}{\textbf{Policy - Employee can access}} \\ \hline
    \textsf{Q1} & \textsf{only her friends' ages}      \\
    \textsf{Q2} & \textsf{all employee details if she is in the HR department; if not, only her own details}                  \\    
    \textsf{Q3} & \textsf{average salary of employees who are not managers unless she is a manager}                             \\
    \textsf{Q4} & \textsf{address of her friends but sees only the city name for other employees} \\ 
    \textsf{Q5} & \textsf{only those events to which she is invited} \\ 
    \textsf{Q6} & \textsf{events at a particular location but sees ``Private event'' for events to which she is not invited} \\ \hline
  \end{tabularx}
  \vspace{2mm}
  \caption{Microbenchmark - Policies enforced for the intranet website}
  \label{tab:queries}
\end{table}

\subsubsection{Performance}
\begin{figure}
  \centering
    \includegraphics[width=\linewidth]{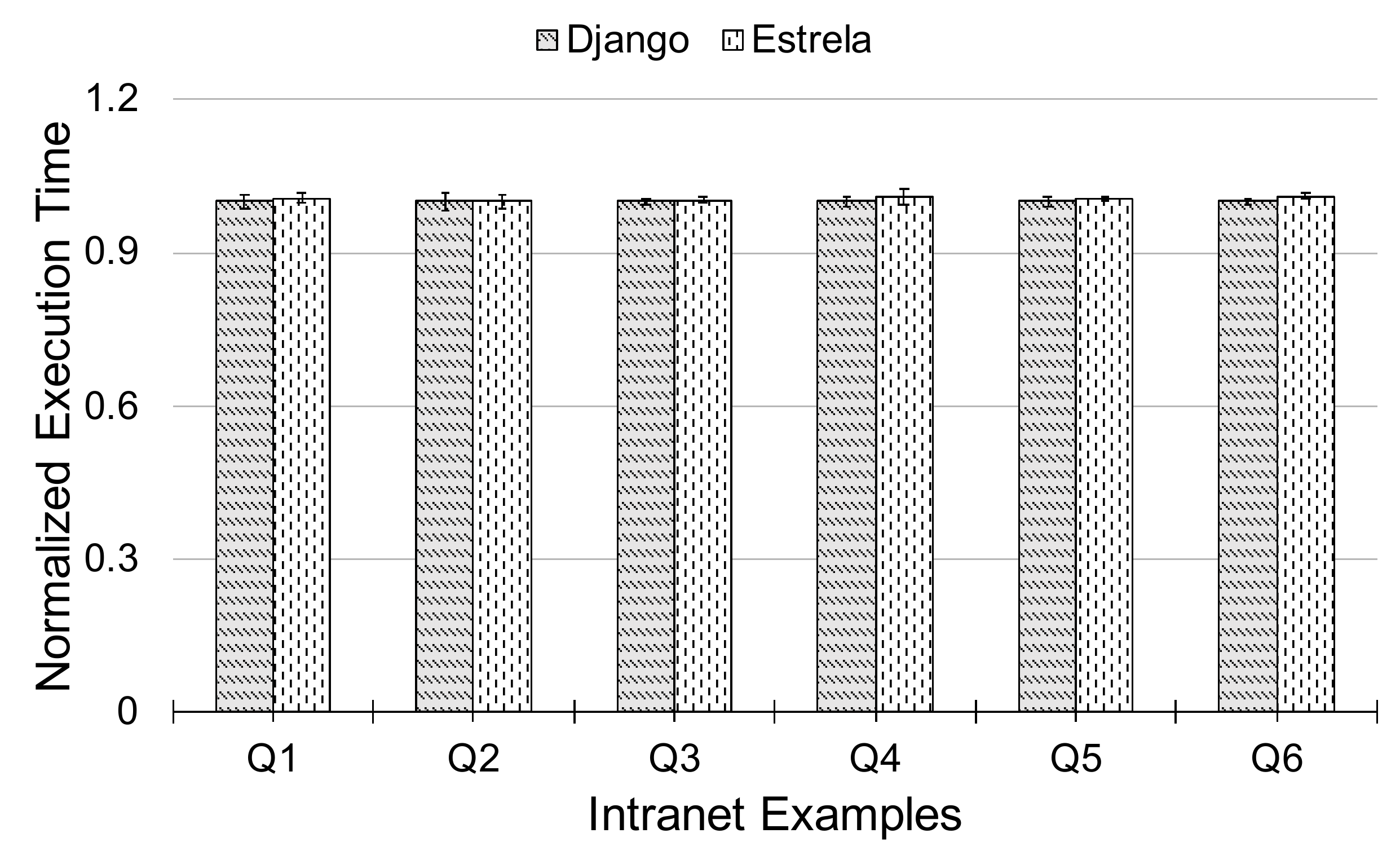}
  \caption{Normalized execution time for intranet examples with Django as the baseline}
  \label{fig:eval-intranet}
\end{figure}
For the intranet example, we added employee details for 20,000 users
in the database with each having at least 5 friends. Additionally, we
added 100000 events to test the scenarios involving events (Q5 and
Q6). The policies for Q4 and Q6 are post-eval policies while for the
other examples are pre-eval. We measure the time taken to access
employee and event information for the six examples shown in
Table~\ref{tab:queries}. The normalized execution time for the
examples with policies enforced in Django and in {\framework} is shown
in Figure~\ref{fig:eval-intranet} with Django being the baseline. The
time includes the time taken to send the request, apply the policy,
execute the query, and to send back the response to the client. The
overhead for policy enforcement in {\framework} ranges from about
$0\%$ when accessing employee details to $1\%$ when accessing a set
of events at a particular location. The additional overhead for
{\framework} is due to the checks to select the correct policies to
apply.  

\paragraph{Throughput}
\begin{figure}
  \centering
    \includegraphics[width=\linewidth]{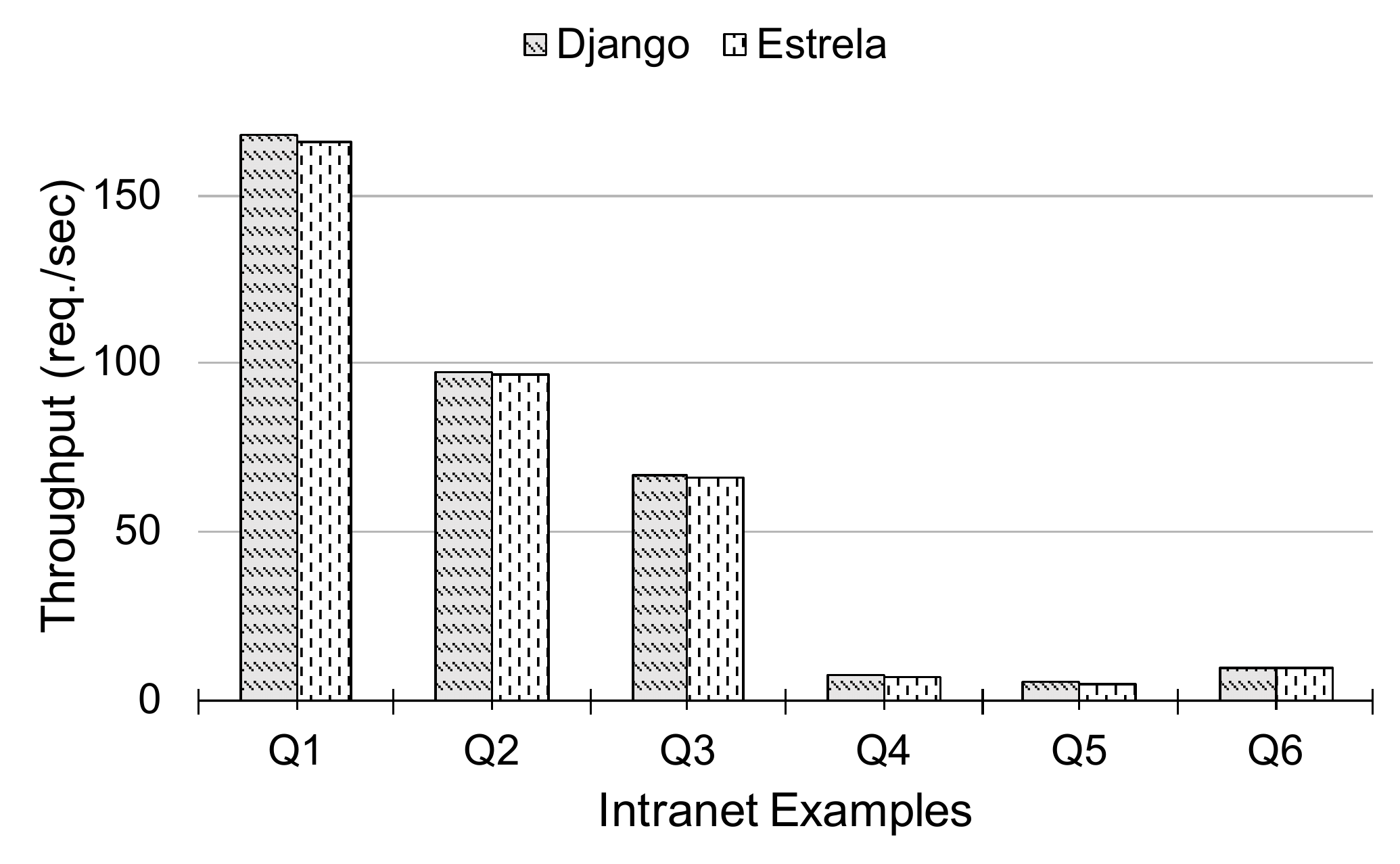}
  \caption{Throughput of {\framework} and Django for the intranet application with a concurrency level of 10}
  \label{fig:throughput}
\end{figure}
To examine {\framework}'s impact on the throughput, we measured the
number of requests handled per second when issuing the queries Q1 - Q6
described above. We used ApacheBench version 2.3 to measure the
throughput with a concurrency level of 10. Each run issued 500
requests. Figure~\ref{fig:throughput} shows the resulting degradation
in throughput, which was around 1\% in the worst-case.   

\subsection{Policy complexity}
\begin{figure}
  \centering
    \includegraphics[width=\linewidth]{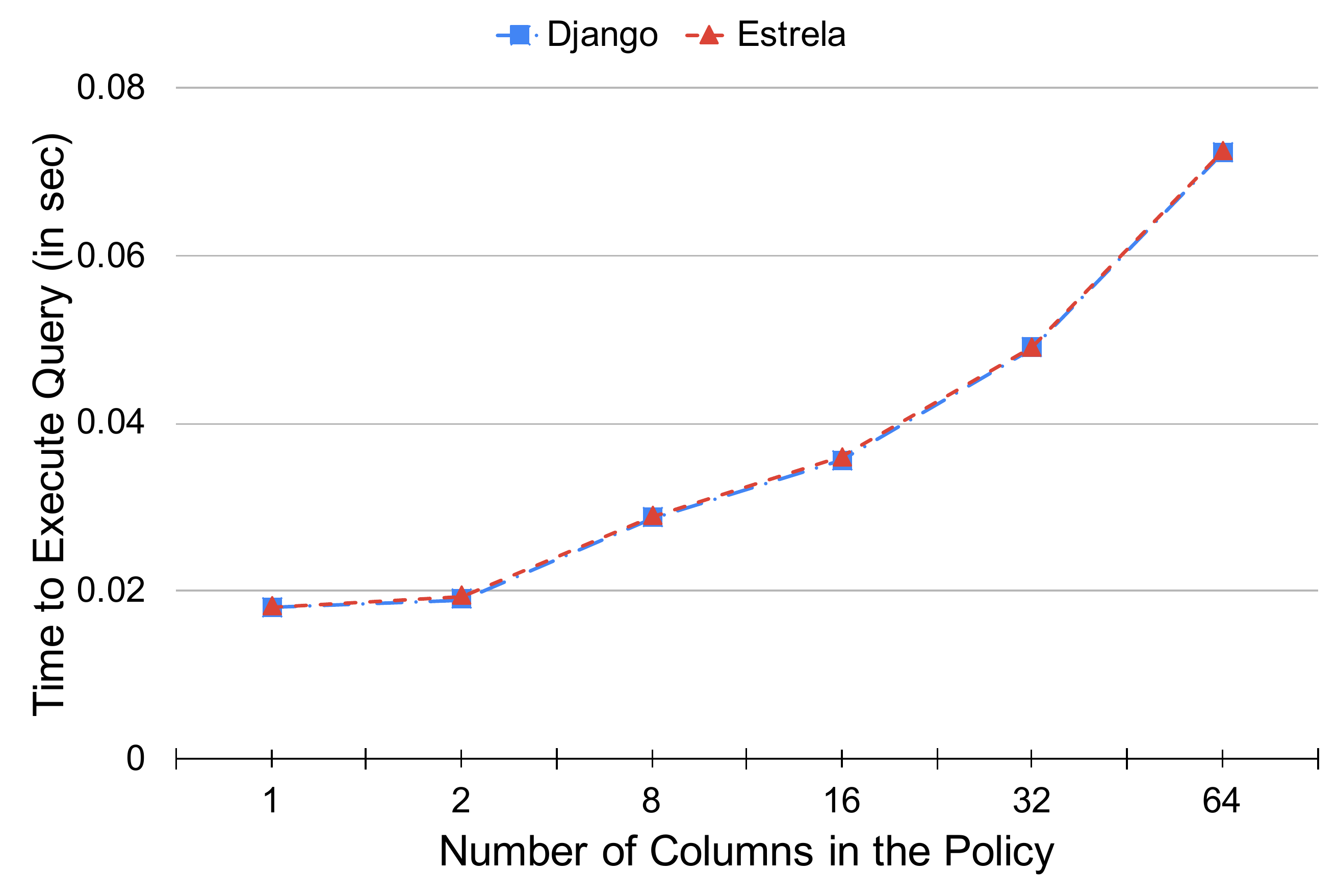}
  \caption{Time taken for executing a query subject to a pre-eval policy with different number of columns}
  \label{fig:complexity}
\end{figure}
We show in Figure~\ref{fig:complexity}, how the complexity of a policy
affects the overhead of {\framework}. We create an application in
{\framework} with a couple of tables each having 100 columns and a
100,000 rows, and test the performance for different number of columns
being used in the policy. We ensure that all conditions in the policy
are evaluated to get the worst-case results, and evaluate it for both
-- Django with policy in the API code, and {\framework}. The overhead
that {\framework} adds is almost constant, and does not increase as
the complexity of the policy increases.  

\section{Related Work}
\label{sec:related}
We describe some of the closely related works to {\framework} on
policy enforcement, most of which deal with enforcing policies on the
server-side of applications that communicate with a database to
retrieve and store sensitive data. 

Qapla~\cite{mehta2017} is a framework to provide fine-grained access
control in database-backed applications where the policies are
specified as SQL WHERE clauses that define what information the users
are allowed to access. Qapla's enforcement engine modifies the
low-level SQL queries made to the database by adding the policies as
sub-queries. However, it does not support the specification of default
values that reveal the existence of a sensitive value, and does not
provide the flexibility of applying release policies except
aggregation offered by the post-eval policies in
{\framework}. Extending Qapla to support post-eval policies is
non-trivial because such policies cannot be specified as SQL WHERE
clauses requiring an additional policy specification
mechanism. Moreover, integrating Qapla with existing applications
might require modifications to the application code to query for the
correct column-transformations without which it returns a more
restrictive set of results.

Hails~\cite{giffin2012} is an \emph{MPVC} web framework that adds
policies declaratively alongside database schemas and tags every piece
of data in the system with a label. These labels are carried around as
the data flows in the system and checked by a trusted runtime when
leaving the system. The focus is to control the flow of data to
untrusted third-party applications by building applications using the
Hails web platform. Similarly, Jacqueline~\cite{yang2016} is a
framework to track information flow in applications dynamically using
the policy-agnostic design paradigm. Jacqueline relies on a modified
database that stores multiple views of data based on who is allowed to
access what. It additionally allows specification of default values,
but does not support policies that are linked to a set of sensitive
fields in the database or policies involving data-aggregates. 
While {\framework} primarily enforces server-side policies without any
client-side information flow tracking, it supports contextual policies
unlike Hails and Jacqueline without requiring modifications to the
database. {\framework} can be integrated with these frameworks for
information flow tracking at the client-side to ensure that the
sensitive data does not flow to unauthorized parties.  


FlowWatcher~\cite{muthukumaran2015} is a system that enforces
information flow policies within a web proxy without requiring
modifications to the application. However, it is difficult to enforce
fine-grained policies, like the ones {\framework} supports, in the
system. LWeb~\cite{parker2019} is another system that provides
information flow control for web applications developed in Haskell,
which supports expressive policies albeit with moderate
overheads. LWeb, however, does not support contextual policies and
requires building all applications from scratch. Similarly,
Daisy~\cite{guarnieri2019} provides flow tracking in databases
supporting features like triggers and dynamic policies. However, their
monitor supports only policies that can be expressed in the SQL access
control model. SELinks~\cite{corcoran2009} allows both server-side and
database-side enforcement of policies by compiling server-side
functions to user-defined functions that can run on the database for
compliance. It does not support contextual policy compliance and is
limited to controlling data disclosure.  

Sen \emph{et al.}~\cite{sen2014} propose Legalease language for
stating policies using Deny and Allow clauses and enforce it on big
data systems. The language is simple allowing easy policy
specification by the developers and does partial information flow
tracking to catch violations of  policies, but can express only a
smaller subset of policies than {\framework}. For instance, Legalease
cannot return parts of sensitive information unless stored  explicitly
in the database and queried explicitly for by the application.  

Ur/Web~\cite{chlipala2015} is a domain-specific language for
programming web applications, which includes a static information flow
analysis called UrFlow~\cite{chlipala2010}. Policies are expressed in
the form of SQL queries and can express what information a particular
user can learn but the analysis might over-approximate being static in
nature. DataLawyer~\cite{upadhyaya2015} is a system to analyze data
usage policies that checks these policies at runtime when a query is
made to the database. The policies are specified in a formal language
based on SQL. It allows quite expressive policies but checks all
policies whenever the database is queried. {\framework}, on the
contrary, selects the right set of policies based on the fields used
in the query, making it more efficient. CLAMP~\cite{parno2009}
protects sensitive data leakage from web servers by isolating user
sessions and instantiating a new virtual web server instance for every
user session. The queries are restricted to data accessible by the
current user; however, the granularity of policies is limited to
per-table.  

Byun and Li~\cite{Byun2008} and later Kabir \emph{et
  al.}~\cite{Kabir2011} proposed purpose-based access control to
enforce purpose-based policies like the ones related to HIPAA showed
in the paper. While purpose-based access control takes contextual
purpose into account, the existing approaches rely on user's
trustworthiness to specify the purpose for access. In contrast, our
work uses path-specific information to determine the purpose
dynamically.  

SIF~\cite{chong2007} is a framework for developing web applications
that respect some confidentiality policies. The framework is built on
top of Jif~\cite{myers1999}, an extension of Java with information
flow control, and enforces information flow control in Java
Servlets. Besides being language-specific, SIF also incurs moderate
overheads because of the analysis. SeLINQ~\cite{schoepe2014} is
another information flow control system to enforce policies across
database boundaries that modifies a subset of F\# with database
queries to perform information flow analysis. IFDB~\cite{schultz2013}
enforces information flow control in databases in a decentralized
fashion by associating labels with every data object. It uses
specialized views for pre-eval policies. RESIN~\cite{yip2009} allows
developers to check data-flows in the program using assertions, which
are checked at sinks in the program without developer-added checks. It
can also express rich policies as assertions in the
program. Nemesis~\cite{dalton2009} focusses on preventing
authentication and access control vulnerabilities in web applications
by using dynamic data flow tracking. These systems label information
and track its flow through the application and guarantee that
information flows as per the policies but are specific to the
respective languages. 

Passe~\cite{blankstein2014} allows enforcement of learned security
policies by running applications as a set of isolated or sandboxed OS
processes and restricting what data each process can access or
modify. The principle behind Passe is to separate the privileges of
various parts of an application as per its requirements and granting
least privilege access, as required, for running the isolated
parts. The constraints are discovered in a learning phase inferring
the data-flows and control-flows in the application to limit the flow
of data. {\framework}, on the other hand, associates
developer-specified policies with APIs. 
 

\section{Conclusion}
\label{sec:conclusion}
We present {\framework}, a framework that ensures policy-compliance in
database-backed applications by supporting specification and 
enforcement of contextual and granular policies. The
policies are specified separately from the application code alongside
the database schema without requiring any modifications to the
existing code and the database. We prototyped {\framework} in Python
on top of Django to specify and enforce API-specific
policies. {\framework} supports easy migration of legacy applications
for policy-compliance. We show the applicability of {\framework} by
building/migrating four applications on top of it and showing that it
incurs low overheads while enforcing expressive policies. 

\begin{acks}
  We would like to thank our shepherd, Qi Li, and the anonymous
  reviewers for their insightful comments and feedback. The work was 
  supported in part by the Center for Machine Learning and Health
  (award no. PO0006063506) at Carnegie Mellon University.   
\end{acks}

\end{document}